\documentclass{emulateapj}
\usepackage{natbib}
\usepackage{epsfig}
\usepackage{graphicx}

\shorttitle{New pulsars in NGC 6440 and NGC 6441}
\shortauthors{Freire {\em et al}}

\begin{document}

\title{Eight New Millisecond Pulsars in NGC~6440 and NGC~6441}

\author{Paulo C. C. Freire\altaffilmark{1},
  Scott M. Ransom\altaffilmark{2},
  Steve B\'egin\altaffilmark{3, 4},
  Ingrid H. Stairs\altaffilmark{3},
  Jason W. T. Hessels\altaffilmark{5},
  Lucille H. Frey\altaffilmark{6},
  and Fernando Camilo\altaffilmark{7}
}

\altaffiltext{1}{National Astronomy and Ionosphere Center, Arecibo
  Observatory, PR 00612; {\tt pfreire@naic.edu}}
\altaffiltext{2}{National Radio Astronomy Observatory,
  Charlottesville, VA 22903}
\altaffiltext{3}{Department of Physics and Astronomy, University of
  British Columbia, Vancouver, BC V6T 1Z1, Canada}
\altaffiltext{4}{D\'epartement de physique, de g\'enie physique et d'optique, Universit\'e Laval, Qu\'ebec, QC G1K 7P4,  Canada}
\altaffiltext{5}{"Astronomical Institute "Anton Pannekoek", University
  of Amsterdam, Kruislaan 403, 1098 SJ Amsterdam, The Netherlands}
\altaffiltext{6}{Department of Astronomy, Case Western Reserve
  University, Cleveland, OH 44106}
\altaffiltext{7}{Columbia Astrophysics Laboratory, Columbia
  University, New York, NY 10027}

\begin{abstract}
  Motivated by the recent discovery of 30 new millisecond pulsars in
  Terzan~5, made using the Green Bank Telescope's S-band receiver and
  the Pulsar Spigot spectrometer, we have set out to use the same
  observing system in a systematic search for pulsars in other
  globular clusters. Here we report on the discovery of five new
  pulsars in NGC~6440 and three in NGC~6441; each
  cluster previously had one known pulsar. Using the most recent
  distance estimates to these clusters, we conclude that there
  are as many potentially observable pulsars in NGC~6440 and NGC~6441
  as in Terzan~5. We present timing solutions for all of the pulsars in
  these globular clusters. Four of the new
  discoveries are in binary systems; one of them,
  PSR~J1748$-$2021B (NGC~6440B), has a wide ($P_b\,=\,20.5\,$d) and
  eccentric ($e\,=\,0.57$) orbit. This allowed a measurement of its
  rate of advance of periastron, $\dot{\omega}\,=\,0.00391(18)^\circ \rm
  yr^{-1}$. If due to the effects of general relativity, the total
  mass of this binary system is $2.92\,\pm\,0.20\,M_{\sun}$ (1 $\sigma$),
  implying a median pulsar mass of $2.74\,\pm\,0.21\,M_{\sun}$. There
  is a 1\% probability that the inclination is low enough that pulsar
  mass is below $2\,M_{\sun}$, and 0.10\% probability that it
  is between 1.20 and 1.44~$M_{\sun}$. If
  confirmed, this anomalously large mass would strongly constrain the
  equation of state for dense matter. The other highly
  eccentric binary, PSR~J1750$-$37A, has $e\,=\,0.71$, and
  $\dot{\omega}\,=\,0.0055(3)^\circ \rm yr^{-1}$, implying a total
  system mass of $1.97\,\pm\,0.15\,M_{\sun}$ and, along with the
  mass function, maximum and median pulsar masses of 1.65 and
  1.26\,$M_{\sun}$ respectively.

\end{abstract}


\section{Introduction}\label{sec:intro}

A recent survey of the globular cluster (GC) Terzan~5 with the S-band
(1.7 to 2.6 GHz) receiver of the Green Bank Telescope (GBT) and the
Pulsar Spigot spectrometer \cite{kel+05} has met with extraordinary
success, finding 30 new
millisecond pulsars (MSPs). The first 21 discoveries were announced in
Ransom et al. (2005)\nocite{rhs+05}. More recently, nine more pulsars
were discovered\footnote{See
  \url{http://www2.naic.edu/$\sim$pfreire/GCpsr.html}}, including
PSR~J1748$-$2446ad, the fastest spinning pulsar known
\cite{hrs+06}. These discoveries show that this observing system
(henceforth GBT/S/PS) has an unprecedented sensitivity to MSPs
outside the Arecibo sky, due in great part to the factor of 3 higher
gain of the GBT as compared to the Parkes telescope.  Perhaps even
more importantly,
the higher observing frequency ($\sim$2\,GHz), large observing
bandwidth ($600$\,MHz), and relatively fine frequency resolution
($\sim$0.78\,MHz) retain high sensitivity and time resolution for MSPs
with large dispersion measures (DM).  The discovery of the
1.39\,ms PSR~J1748$-$2446ad at a DM of 235.6 cm$^{-3}$pc dramatically
demonstrates the sensitivity of the GBT/S/PS system to fast-spinning
pulsars at high DMs. This motivated us to search for MSPs in other
promising GCs with the GBT/S/PS system.

\begin{deluxetable*}{ l r c r c c r r c}
\scriptsize
\tablecolumns{8}
\tablewidth{0pc}
\tablecaption{The six globular clusters with highest $\Gamma_c$}
\tablehead{
\colhead{Globular} &
\colhead{$D$\tablenotemark{a}} &
\colhead{$\rho_0$\tablenotemark{b}} &
\colhead{$\Gamma_c$} &
\colhead{DM\tablenotemark{c}} &
\colhead{$\tau_{\rm scatt, 2 GHz}$\tablenotemark{d}} &
\colhead{$N_{\rm psr}$\tablenotemark{e}} &
\colhead{$\bigtriangleup T$\tablenotemark{f}} &
\colhead{DM Ref.}
\\
\colhead{Cluster} &
\colhead{(kpc)} &
\colhead{(log [L$_{\sun}$\,pc$^{-3}$])} &
\colhead{(\%)} &
\colhead{(cm$^{-3}$pc)} &
\colhead{($\mu$s)} &
\colhead{} &
\colhead{(hr)} &
\colhead{\tablenotemark{g}}
}
\startdata
 NGC 6388 &    10.0   &  5.34  &  9.5 & 318 & 1.4   & 0  &
 2.2 & \\
 Terzan 5 &     10.3  &  5.06  &  8.6 & (234.3--245.6) & 6
 & 3 $+${\bf 30} & 8.0 & (1)\\
 NGC 6441 &    11.7   &  5.25  &  8.0 & (230.1--234.4) &
 2  & 1 $+${\bf 3}  & 5.3 & (2)\\
 NGC 6440 &     8.4   &  5.28  &  6.4 & (219.4--227.0) & 5
 & 1 $+${\bf 5}  & 8.4 & (3)\\
 NGC 6266 &      6.9  &  5.14  &  5.1 & (113.4--115.0) &
 4  & 6  &   7.0 & (4)\\
 Liller 1 &      9.6  &  5.53  &  4.2 & 771 & $\sim 3 \times 10^2$ & 0  &
 6.2  & \\
\enddata
\tablenotetext{a}{Distance from the Solar System from Harris (1996).
  These values have significant systematic uncertainties.}
\label{tab:clusters}
\tablenotetext{b}{Central density of GC; see last column of third
  table in \url{http://physwww.mcmaster.ca/$\sim$harris/mwgc.dat}.}
\tablenotetext{c}{For GCs with known pulsars, we present the
DM range for the known pulsars, with references in the last column,
otherwise we predict the DM for the GC's estimated distance using
  the Cordes \& Lazio (2001) model. These can be off by a factor of up to about 2.}
\tablenotetext{d}{Extrapolated from other frequencies using $\tau_{\rm
  scatt, 2 GHz} \, = \, \tau_{\rm scatt}(F) (F / {\rm 2
  GHz})^{4.4}$. In the case of Terzan 5, the scattering was measured
  by Nice \& Thorsett (1992)\nocite{nt92} as 700\,$\mu$s at
  685\,MHz. Otherwise, we scale the scattering time at 1GHz predicted
  by the Cordes \& Lazio (2001) model for the DM in the previous
  column. These can be off by up to an order of magnitude.}
\tablenotetext{e}{Number of previously known pulsars $+$ {\bf number of new
  pulsars}.}
\tablenotetext{f}{Amount of time the GC is visible at the GBT
  per day}
\tablenotetext{g}{(1) Ransom et al. 2005\nocite{rhs+05}, (2) Possenti
  et al. 2006\nocite{pcm+06}, this work, (3) Lyne, Manchester \&
  D'Amico 1996, this work, (4) Possenti et al. 2003\nocite{pdm+03},
  Chandler 2003.}
\end{deluxetable*}

\begin{deluxetable*}{ l c c c}
\scriptsize
\tablecolumns{4}
\tablewidth{0pc}
\tablecaption{Parameters for three globular clusters}
\tablehead{ \colhead{ } & \colhead{NGC 6388} & \colhead{NGC 6440}
& \colhead{NGC 6441}}
\startdata
Right Ascension of center, $\alpha$ \dotfill &
$17^{\rm h}36^{\rm m}17\fs 0$ &
$17^{\rm h}48^{\rm m}52\fs 7$ &
$17^{\rm h}50^{\rm m}12\fs 9$ \\
Declination of center, $\delta$ \dotfill &
$-44^\circ 44\arcmin 06\arcsec$ &
$-20^\circ 21\arcmin 37\arcsec$ &
$-37^\circ 03\arcmin 05\arcsec$ \\
Galactic longitude, $l$ ($^\circ$) \dotfill & 345.56 & 7.73 & 353.53 \\
Galactic latitude, $b$ ($^\circ$) \dotfill & $-$6.74 & 3.80 & $-$5.01 \\
Cluster distance, $D$ (kpc) \dotfill & 10.0 & 8.2\tablenotemark{a} & 13.5\tablenotemark{a} \\
Angular core radius, $\theta_c$ (arcmin) \dotfill & 0.12 & 0.13 & 0.11 \\
Half-mass radius, $\theta_h$ (arcmin) \dotfill & 0.67 & 0.58 & 0.64 \\
Metallicity ($\log$[Fe/H]) \dotfill & $-0.60$ & $-0.34$ & $-0.53$ \\
$v_z(0)$ (km s$^{-1}$) \dotfill & 19.13\tablenotemark{b} & 13.01\tablenotemark{b} & $19.5^{+9.4}_{-5.1}$\tablenotemark{c} \\
$a_G$\tablenotemark{d} ($10^{-9}$m\,s$^{-2}$) \dotfill & $-$0.46 & $-$0.029 & $-$0.46 \\
\enddata
\tablecomments{Parameters as in Harris (1996), except where
  indicated.}
\tablenotetext{a}{Valenti et al. (2007).\nocite{vfo07}}
\tablenotetext{b}{Webbink (1985).\nocite{web85}}
\tablenotetext{c}{Dubath et al. (1997).\nocite{dmm97}}
\tablenotetext{d}{Calculated using the data above and the Kuijken \&
  Gilmore (1989) model of the Galaxy.}
\label{tab:GC}
\end{deluxetable*}

\subsection{Cluster selection}

In the high stellar density environments in the cores of GCs, binary
systems with at least one main sequence (MS) star are occasionally
disrupted by the intrusion of a neutron star (NS).  The most likely
outcome is an ``exchange encounter'', where the lightest component of
the previous binary is ejected and a NS-MS binary remains.  The
resulting NS-MS systems can become much more numerous than those
formed from evolution of primordial binaries.  With the evolution of
the MS star, it can happen that matter will accrete onto the NS during
a Low-Mass X-ray Binary (LMXB) phase, particularly among the more
compact systems. When the LMXB accretion stops, we are left with an MSP
orbiting a white dwarf (WD) star.

Pooley et al.~(2003)\nocite{pla+03} showed a strong correlation
between the number of X-ray point sources in a GC (which are generally
compact binaries) and the rate of
stellar encounters within the half-mass radius, a parameter that they
designate $\Gamma$. This correlation is stronger than for any other
individual GC parameter, and applies to active LMXBs as well (although
with only a dozen systems in the Galactic GC system, the statistics
are poorer). $\Gamma$ might therefore be a good indicator of the
present rate of MSP formation. Pooley et al.~(2003) provide a simple
way of calculating the stellar interaction rate at the core of any GC:
\begin{equation}
\label{eq:gamma}
\Gamma_c \propto \rho_0^{1.5} r_c^2,
\end{equation}
where $\rho_0$ is the central mass density and $r_c$ is the core
radius. This parameter is perhaps more relevant than $\Gamma$ for the
formation of LMXBs because mass segregation constrains neutron stars
to within a few core radii of the center of the GC.

We calculated $\Gamma_c$ for all Galactic GCs with the central
luminosity densities and core radii given in
Harris~(1996)\nocite{har96}\footnote{See
  http://physwww.mcmaster.ca/$\sim$harris/mwgc.dat for an updated list
  of GC parameters.}, and assuming a constant conversion factor
between central luminosity density and central mass density. The six
best GCs are listed in Table \ref{tab:clusters}, with their $\Gamma_c$
presented as a percentage of the total $\Gamma_c$ summed over all
Galactic GCs. This table represents our initial survey expectations;
we hypothesized that the percentages should be closely related to the
fraction of the total GC MSP population contained in each GC.

For the first phase of the survey, we selected the GCs in Table
\ref{tab:clusters} that had not been previously searched with the
GBT/S/PS system: NGC~6388, NGC~6440 and NGC~6441.  Liller~1 was
searched at the same time as Terzan~5 by Ransom et al.~(2005) and
NGC~6626 has been searched by Jacoby et al.~(2002)\nocite{jcb+02}.
Despite the large predicted stellar interaction rates, only two
pulsars were known in these GCs: PSR~B1745$-$20A (NGC~6440A; Lyne
et al.~1996)\nocite{lmd96} and PSR~J1750$-$37A
(NGC~6441A; Possenti et al.~2006)\nocite{pcm+06}. The spin periods and
DMs of NGC~6440A and NGC~6441A are 288.6\,ms and 111.6\,ms and
220 and 234~cm$^{-3}$\,pc, respectively.  We assumed that
the lack of other known pulsars in these high-DM GCs was largely due to
a bias of the earlier surveys against the detection of faster-spinning MSPs.
The high central frequency and improved time resolution of the
GBT/S/PS system greatly diminish pulse broadening due to multi-path
propagation and dispersive smearing, and would therefore alleviate the
bias against the detection of other MSPs in these GCs.

\subsection{Data taking}
\label{sec:datataking}

From 2005 April 18 to August 21, we observed NGC~6388, NGC~6440 and
NGC~6441 on three different occasions each, using the GBT's S-band
receiver with the Pulsar Spigot. The longest integration times used
for each GC are very close to their maximum track times indicated
in Table~\ref{tab:GC}.

These observations take advantage of the relatively
radio-frequency-interference (RFI) clean band between 1650 and
2250\,MHz. The Pulsar Spigot is built around the GBT correlation
spectrometer and is available in a number of modes \cite{kel+05}. For
most observations, we used the Spigot in mode 2 where two orthogonal
polarizations with a total bandwidth of 800\,MHz are digitized with
three levels, autocorrelated with 1024 lags, and integrated for
81.92\,$\mu$s. The resulting autocorrelation functions are summed and
then written to disk as 16-bit integers. After Fourier transforming
the lags, we have 768 usable channels across the 600-MHz bandpass
(0.78125-MHz channels). At a frequency of 2\,GHz and a DM of
220~cm$^{-3}$\,pc, the dispersive smearing across one channel is
178\,$\mu$s. Recently, we have been using mode 14, this mode writes
the 81.92-$\mu$s integrations to disk as 8-bit integers, but with
twice the frequency resolution, resulting in a dispersive smearing
across one channel for a DM of 220~cm$^{-3}$\,pc of 89\,$\mu$s.

In the case of NGC~6440 and NGC~6441, the discovery of new pulsars
lead us to observe these two GCs on a total of 37 and 24 extra
occasions respectively, from 2005 November 5 to 2007 March 28 (Spigot
mode 2), and then from 2007 October 10 to November 1 (Spigot mode 14),
the latter data has only been added to the ephemeris of NGC~6440B (see
\S \ref{sec:timing}).  The typical observing session spends the first
2-3 hours on NGC~6440, then observes NGC~6441 for 2.5 to 4 hours, and
then spends 2 to 3 hours again on NGC~6440. The total usable observing
times are about 90--100\,hr for each pulsar in NGC~6440 and NGC~6441.

On two occasions, we observed NGC~6440 and NGC~6441
at 820\,MHz with a bandwidth of 50\,MHz in order to determine the DMs
of the pulsars more accurately and to search for pulsars with spectral
indices too steep to be detected at 1950\,MHz, or that lie outside the
area covered by the S-band beam.
Because the telescope beam widths at these frequencies (6\farcm5 at
1.95 GHz and 15\arcmin \ at 820~MHz) are significantly larger than the
half-mass radii of these GCs (see Table~\ref{tab:GC}), we
observed only single positions centered on the GCs.

\begin{figure*}[htp]
  \begin{center}
    \includegraphics[width=7in,angle=0]{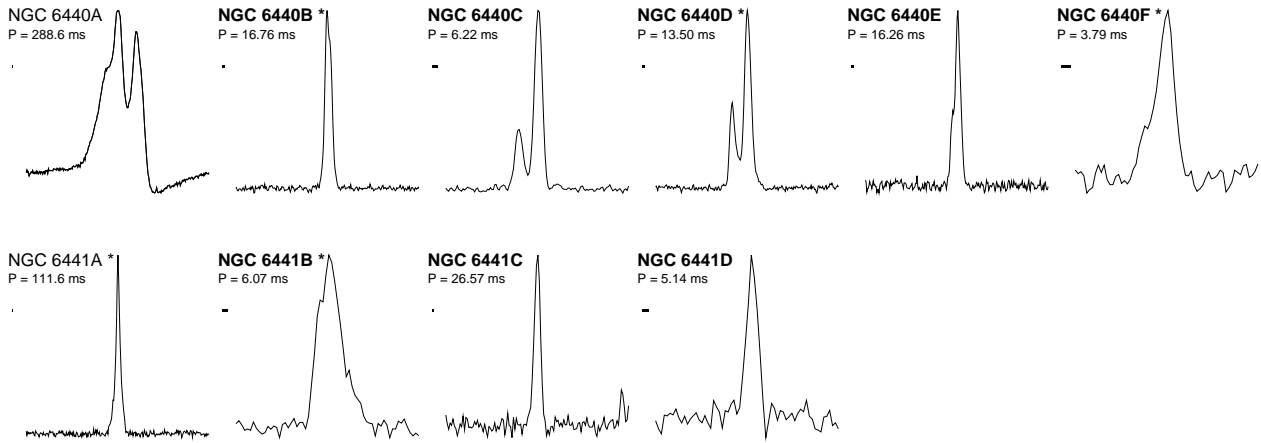}  
    \figcaption{Average 1950 MHz pulse profiles for the 10 pulsars
      known in the GCs NGC~6440 and NGC~6441, obtained by adding all
      observations made with mode 2 of the Pulsar Spigot. These
      profiles cover one full rotation. The pulsars in binary systems
      have an asterisk after their names, and the newly discovered
      pulsars have their names in boldface. The horizontal width of
      the rectangles indicates the system's total time resolution in
      mode 2, including the effects of dispersive smearing, relative
      to each pulsar's spin period. The dip in power after the main
      pulse of NGC~6440A is likely a Spigot artifact.
    \label{fig:profiles}}
  \end{center}
\end{figure*}

\section{Search and discovery of new pulsars}

After masking the interference (in the manner described in detail in
Hessels et al.~2007), the search data (taken in Spigot mode 2) were
dedispersed in 48 sub-bands over the whole 600-MHz usable band at the
``average'' DMs for NGC~6440 (223\,cm$^{-3}$\,pc) and NGC~6441
(233\,cm$^{-3}$\,pc), reducing the volume of data to be stored by a
factor of $\sim$20. For each sub-band set we created $\sim$40
dedispersed time series covering a range of trial DMs
$\pm$10\,cm$^{-3}$\,pc around the nominal GC DMs in steps of
0.5\,cm$^{-3}$\,pc. If a pulsar has a DM halfway between two trial
DMs, that introduces an extra smearing of 168$\mu$s, resulting in a
total maximum smearing of 245\,$\mu$s. We also barycentered each time
series using {\sc
  TEMPO}\footnote{http://www.atnf.csiro.au/research/pulsar/tempo/}.

We Fourier transformed each time series and searched them using a
frequency domain acceleration search technique (Ransom et al.~2002)
\nocite{rem02} that allows a fully coherent search in
frequency/frequency-derivative space as well as the incoherent
addition of a number of harmonics (1, 2, 4, 8 or 16). The technique
has an optimal sensitivity for pulsars in binary systems with an
orbital period much longer than the observation time (P$_{\rm orb}
\geq 10$T$_{\rm obs}$). For this reason, the acceleration
searches were performed on all the observations for the entire length
of each observation as well as on overlapping segments of various
shorter durations (typically 27 and 82 minutes). A more detailed
description of the search procedure is presented in Hessels~et~al.~(2007).

We discovered eight new MSPs, all in S-band data, five of them in
NGC~6440 and three in NGC~6441. Preliminary results on these objects
were presented in B\'egin (2006)\nocite{beg06}. Their pulse profiles,
together with those of the previously known objects, are presented in
Figure~\ref{fig:profiles}. These pulse profiles were obtained by
adding {\em all} detections of the pulsars made in mode 2, carefully
excluding all sub-integrations in time or frequency that are corrupted
by significant RFI.

We found no pulsars in NGC~6388, possibly due to
dispersive smearing and/or interstellar scattering being substantially
higher than the prediction given in Table~\ref{tab:clusters}. The
search for pulsars in NGC~6388 was further hindered by the small time
the GBT can observe the GC (2.2\,hrs; see Table
\ref{tab:clusters}). Furthermore, many of the pulsars in Terzan~5,
NGC~6440 and NGC~6441 were found by searching follow-up
timing observations, none of which were made for NGC~6388.

\begin{figure}[htp]
  \begin{center}
    \includegraphics[width=3in,angle=0]{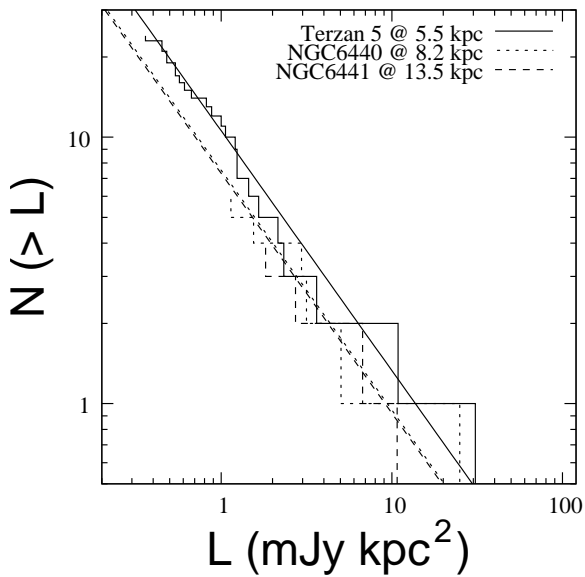}  
    \figcaption{Cumulative distribution of pseudo-luminosities at
    1950 MHz for the pulsars in NGC~6440, NGC~6441 and Terzan 5. The
    pseudo-luminosities are estimated assuming the most recent GC
    distance estimates (Ortolani et al. 2007, Valenti et al. 2007).
    \label{fig:fluxes}}
  \end{center}
\end{figure}

\subsection{Flux densities}
\label{sec:numbers} 

The 1950~MHz flux densities ($S_{1950}$) of all the pulsars in
NGC~6440 and NGC~6441 and their pseudo-luminosities at the same
frequency ($L_{1950}\equiv S_{1950} D^2$) are listed in
Tables~\ref{tab:isolateds} (isolated pulsars) and \ref{tab:binaries}
(binary pulsars). These were determined from the pulse profiles in
Fig.~\ref{fig:profiles} by assuming that the off-pulse
r.m.s. is as predicted by the radiometer equation for the total
observing time. In calculating the system's equivalent flux
density, we used a system temperature of 26K towards
NGC~6440 and 28 K towards NGC~6441, a gain of 1.95 K/Jy for the GBT at
S-band and multiplied the result by 1.235 for the 3-level efficiency
factor of the correlator.

These flux densities indicate that some of
the new discoveries, particularly in NGC~6441, are as faint as the
faintest pulsars in Terzan~5, suggesting that we have reached a
similar sensitivity in our survey as in that reported by
Ransom~et~al.~(2005).

We can compare the sizes of the pulsar populations in these clusters
with that of Terzan 5 by comparing the distribution of their
pseudo-luminosities. The observing system and observing frequency are
the same, the DMs of the pulsars are very similar, the scattering
times are similarly small, the length of the integrations
are similar (Table \ref{tab:clusters}) and the search software is the
same, making the search similarly sensitive to fast-spinning pulsars
(\S \ref{sec:periods}). Furthermore, the software used to make the flux
density measurements is similar (but not exactly - some refinements
have recently been added to the program). All these factors make the
flux densities directly comparable; the pseudo-luminosities will then
be directly comparable as well if the GC distances are accurate.

We have fitted a luminosity function of the type $N(> L) = k_1
L^{-0.9}$ (derived by Hessels et al. 2007 as a best fit to the
luminosities of 41 isolated pulsars in M5, M13, M15, M28, Terzan 5, 47
Tucanae, NGC~6440 and NGC~6441 above L$_{1400}\,=\,1.5$ mJy~kpc$^2$),
where $N(> L)$ is the number of pulsars above a pseudo-luminosity of
$L$ and $k_1$ is the number of pulsars brighter than 1 mJy kpc$^2$.
If we assume the distances listed in Harris~(1996, see Table
\ref{tab:clusters}), we obtain $k_1 = 32.8, 7.9$ and $5.7$ for
Terzan~5, NGC~6440 and NGC~6441 respectively. This would indicate
that $\Gamma_c$ does a poor job at predicting the number of MSPs in
these GCs. 

However, recent studies of these GCs have changed these
distance estimates (see Table \ref{tab:GC}), sometimes drastically: in
the case of Terzan 5, the most recent distance estimate is 5.5 kpc
\cite{obb+07}. It should be noted that these new distance estimates
still have large systematic uncertainties (Heinke and Ortolani,
private communication). These values affect our prediction for
$\Gamma_c$: the total absolute visible luminosity of the core of the GC is
proportional to $D^2$, but the volume of the core is proportional to
$D^3$, therefore $\rho_0 \propto D^2 / D^3 = D^{-1}$. Inserting this
in Eq.  \ref{eq:gamma}, and keeping in mind that $r_c$ is also
proportional to $D$, we obtain $\Gamma_c \propto D^{1/2}$. With the
revised distance estimates, we obtain revised $\Gamma_c$ values of
6.3, 6.3 and 8.6\% for Terzan~5, NGC~6440 and NGC~6441 respectively.

When we use the new distance estimates to re-calculate the
pseudo-luminosities of the pulsars in these three GCs
(Fig.~\ref{fig:fluxes}), we obtain $k_1 = 10.6, 7.5$ and $7.3$ for
Terzan~5, NGC~6440 and NGC~6441 respectively. Given the uncertainties
in the distance estimates, this is consistent with all the clusters
having similarly sized pulsar populations, and $\Gamma_c$ being a good
predictor of the number of pulsars in a GC.

If the $k_1$ numbers reflect the sizes of the pulsar populations,
then assuming the Hessels et al. (2007) luminosity law, we should
detect
\begin{equation}
 N_{\rm PSR, NGC6440} = N_{\rm PSR, Ter5} (k_{1, \rm NGC6440}/k_{1, \rm
Ter 5}) (D^2_{\rm Ter5} / D^2_{\rm NGC6440})^{0.9}\,=\,10.4
\end{equation}
pulsars in NGC~6440 and 4.1 pulsars for NGC~6441\footnote{The
number of pulsars in Terzan 5 we use in this calculation is 30. This
many were detected using the same search methods employed in the
present survey.}. For NGC~6441, we
detect as many pulsars as expected from the luminosity distribution
given its greater distance, but we don't detect faint (10 - 17
$\mu$Jy) pulsars in NGC~6440. This could happen if some of them are in
tight binary systems, which would make them more difficult to
detect. In any case, the most likely reason why we detect more pulsars
in Terzan~5 is simply because that GC is closer to us.

To summarize, we can see from the discussion above that the distance
to the GC does not have a major impact on the estimate of
$\Gamma_c$ ($\Gamma_c \propto D^{1/2}$), but it has a fundamental
impact in estimating pulsar populations from the observed flux
densities ($k_1 \propto D^{2 \times 0.9}$). For Terzan~5, NGC~6440 and
NGC~6441, the use of the latest distance estimates makes $\Gamma_c$
a good predictor of the relative sizes of the pulsar populations.

\subsection{Spin period distribution}
\label{sec:periods}

Looking at the spin periods in Table \ref{tab:limits}, and comparing
them with the spin periods of the pulsars in other GCs\footnote{See
  http://www.naic.edu/$\sim$pfreire/GCpsr.html.}, we notice a
conspicuous absence of fast-spinning objects.  Of the 33
known pulsars in Terzan 5 (a GC for which the discovery
observations have very nearly the same time resolution as in our
survey), 15 have spin periods shorter than that of NGC~6440F, the fastest
MSP we found, and 23 have spin periods shorter than that of NGC~6441D.

This could be a result of interstellar scattering being worse for
NGC~6440 and NGC~6441 than it is for Terzan 5. Looking at
Table~\ref{tab:clusters}, we can see that both NGC~6440 and NGC~6441
have predicted scattering times {\em lower} than those of Terzan~5.
An inspection of the pulse profiles in Figure~\ref{fig:profiles},
particularly of the shorter-period pulsars such as NGC~6440C and
NGC~6441D, does not reveal the typical signature of interstellar
scattering, an exponential decay of intensity after the main peak, as
one would expect from the small scattering timescales in
Table~\ref{tab:clusters}. The difference, if statistically
significant, is likely to be real, not due to scattering or dispersive
smearing.

A two-sided Kolmogorov-Smirnov (2K-S) test \cite{ptvf92} gives an 87\%
probability that the spin period distributions NGC~6440 and NGC~6441
are those of a common population. Comparing these two with
Terzan~5, we obtain, respectively, 1.1\% and 1.8\% probabilities for
the same hypothesis. Comparing these two with the pulsar population in
47~Tucanae, we obtain probabilities of 0.13\% and 0.27\% respectively.

This phenomenon is not restricted to NGC~6440 and NGC~6441.
A recent 1.4\,GHz GC pulsar survey made with the Arecibo telescope
(Hessels et al. 2007)\nocite{hrs+07} has searched for pulsars in
M15. Despite being even more sensitive to short-period pulsars than
the Terzan 5 survey, it found none. The spin period distribution of
the M15 pulsars detected by Hessels~(2007) is indistinguishable from
those of NGC~6440 and NGC~6441 (2K-S probabilities of 62\% and 65\%),
but different from that of Terzan~5 (2K-S probability of 2.6\%).

\subsection{Dispersion measure distribution}
\label{sec:DMs}

The DMs of the known pulsars in NGC~6440 range from 219.4 to
226.9\,cm$^{-3}$\,pc, a span of 7.5\,cm$^{-3}$\,pc. In Terzan~5, the
known pulsars have DMs between 234.3 to 243.5~cm$^{-3}$pc, a span of
9.2\,cm$^{-3}$\,pc \cite{rhs+05}. However, two of the pulsars in
Terzan 5 (A and D) have anomalously high DMs.
Each of these pulsars is spatially more distant
from the center of the GC than the other pulsars. All the other
pulsars have DMs between 234.32 and 239.82\,cm$^{-3}$\,pc. If we
randomly choose six pulsars in Terzan 5, there is a 66\% probability
that the set will include none of the two outliers. In that case, the
maximum DM span will be 5.5\,cm$^{-3}$\,pc, lower than but comparable
to what we observe in NGC~6440.

This is consistent with the idea that the variations in DM between
pulsars are generally due to differences in the Galactic electron
column density\footnote{The exception is 47~Tucanae \cite{fkl+01},
  where the DM differences between pulsars are predominantly due to
  gas intrinsic to the GC.}. Since both NGC~6440 and Terzan~5 are
at similar DMs, and have their pulsars spread over roughly similar
areas of the sky, the differences in DM between the pulsars should be
similar. The spread in DMs seems to be linearly correlated with the
absolute DM of the GC \cite{fhn+05}, an idea which is supported
by the observed spread of DMs in NGC~6440.

For NGC~6441, where only four pulsars are known, the DMs range between
230.09 and 234.39\,cm$^{-3}$\,pc. The slightly more compact
distribution of the smaller number of pulsars is likely to produce a
smaller range of DMs than observed in NGC~6440.

\begin{deluxetable*}{ l c c c c c}
\footnotesize
\tablecolumns{6}
\tablewidth{0pc}
\tablecaption{Parameters for the isolated pulsars in NGC 6440 and NGC 6441}
\tablehead{ \colhead{ }
& \colhead{PSR~B1745$-$20A}
& \colhead{PSR~J1748$-$2021C}
& \colhead{PSR~J1748$-$2021E}
& \colhead{PSR~J1750$-$3703C}
& \colhead{PSR~J1750$-$3703D}\\
\colhead{ } 
& \colhead{(NGC~6440A)}
& \colhead{(NGC~6440C)}
& \colhead{(NGC~6440E)}
& \colhead{(NGC~6441C)}
& \colhead{(NGC~6441D)}
}
\startdata
\multicolumn{6}{l}{Observation and data reduction parameters} \\
\hline
$S_{1950}$\tablenotemark{a} (mJy) \dotfill & 0.37 & 0.044 & 0.023 & 0.015 & 0.010 \\
$L_{1950}$\tablenotemark{b} (mJy kpc$^2$) \dotfill & 25 & 3.0 & 1.5 & 2.7 & 1.8 \\
Number of TOAs \dotfill & 107 & 161 & 61 & 82 & 68 \\
Residual rms ($\mu$s) \dotfill & 452 & 38 & 21 & 100 & 53 \\
EFAC \dotfill & 1.72 & 3.37 & 1.02 & 1.17 & 1.25 \\
Reference Epoch (MJD) \dotfill & 54000 & 54000 & 54000 & 54000 & 54000 \\
\hline
\multicolumn{6}{l}{Timing parameters} \\
\hline
$\alpha$ (J2000) \dotfill &
$17^{\rm h}48^{\rm m}52\fs 689(2)$ &
$17^{\rm h}48^{\rm m}51\fs 17320(15)$ &
$17^{\rm h}48^{\rm m}52\fs 80040(14)$ &
$17^{\rm h}50^{\rm m}13\fs 4541(7)$ &
$17^{\rm h}50^{\rm m}13\fs 0969(4)$ \\
$\delta$ (J2000) \dotfill &
$-20^\circ 21\arcmin 39\farcs7(5)$ &
$-20^\circ 21\arcmin 53\farcs81(4)$ &
$-20^\circ 21\arcmin 29\farcs38(3)$ &
$-37^\circ 03\arcmin 05\farcs 58(3)$ &
$-37^\circ 03\arcmin 06\farcs 368(17)$ \\
$\nu$ (Hz) \dotfill &
  3.464969947246(19) &
160.59270990836(6) &
 61.48547652886(2) &
 37.63830424257(6) & 
194.55492232569(15) \\
$\dot{\nu}$ (10$^{-15}$ Hz s$^{-1}$) \dotfill &
$-$4.7944(12) &
1.543(4) &
$-$1.1812(15) &
1.411(4) &
$-$18.652(9)\\
DM ($\rm cm^{-3}\,pc$) \dotfill & 
219.4(2) & 
226.95(6) & 
224.10(4) & 
230.67(2) & 
230.09(17) \\
\enddata
\tablenotetext{a}{The uncertainties in these fluxes are of the order
  of 10\%.}
\tablenotetext{b}{In this and the next table, the distances used to
  calculate this parameter are those given in Table~\ref{tab:GC}}
\label{tab:isolateds}
\end{deluxetable*}

\begin{deluxetable*}{ l c c c c c}
\footnotesize
\tablecolumns{6}
\tablewidth{0pc}
\tablecaption{Parameters for the binary pulsars in NGC 6440 and NGC 6441}
\tablehead{
\colhead{ } 
& \colhead{PSR~J1748$-$2021B}
& \colhead{PSR~J1748$-$2021D}
& \colhead{PSR~J1748$-$2021F}
& \colhead{PSR~J1750$-$37A}
& \colhead{PSR~J1750$-$3703B}\\
\colhead{ } 
& \colhead{(NGC~6440B)}
& \colhead{(NGC~6440D)}
& \colhead{(NGC~6440F)}
& \colhead{(NGC~6441A)}
& \colhead{(NGC~6441B)}
}
\startdata
\multicolumn{6}{l}{Observation and data reduction parameters} \\
\hline
$S_{1950}$ (mJy) \dotfill & 0.047 & 0.075 & 0.017 & 0.059 & 0.037 \\
$L_{1950}$ (mJy kpc$^2$) \dotfill & 3.2 & 5.0 & 1.1 & 10.8 & 6.7 \\
Number of TOAs \dotfill & 1051 & 946 & 52 & 129 & 277 \\
Residual rms ($\mu$s) \dotfill & 46 & 42 & 36 & 86 & 116 \\
EFAC \dotfill & 1.20 & 1.33 & 1.35 & 1.08 & 1.55 \\
Reference Epoch (MJD) \dotfill & 54000 & 54000 & 54000 & 54000 & 54000 \\
\hline
\multicolumn{6}{l}{Timing parameters} \\
\hline
$\alpha$ (J2000) \dotfill &
$17^{\rm h}48^{\rm m} 52\fs 95290(8)$ & 
$17^{\rm h}48^{\rm m} 51\fs 64665(7)$ & 
$17^{\rm h}48^{\rm m} 52\fs 3339(3)$ & 
$17^{\rm h}50^{\rm m} 13\fs 8016(5)$ & 
$17^{\rm h}50^{\rm m} 12\fs 1770(4)$\\
$\delta$ (J2000) \dotfill & 
$-20^\circ 21\arcmin 38\farcs86(19)$ &
$-20^\circ 21\arcmin 07\farcs414(18)$ & 
$-20^\circ 21\arcmin 39\farcs33(9)$ &
$-37^\circ 03\arcmin 10\farcs95(2)$ & 
$-37^\circ 03\arcmin 22\farcs93(2)$ \\
$\nu$ (Hz) \dotfill & 
59.665418222544(17) & 
74.097014488305(13) & 
263.5998326931(2) & 
8.96050624775(3) & 
164.62146212170(16) \\
$\dot{\nu}$ (10$^{-15}$ Hz s$^{-1}$) \dotfill & 
1.1717(6) & 
$-$3.2216(9) & 
0.733(13) & 
$-$0.4545(6) & 
$-$0.520(11)\\
DM ($\rm cm^{-3}\,pc$) \dotfill & 220.922(11) & 224.98(3) & 220.43(8) & 233.82(3) & 234.391(9) \\
$P_b$ (d) \dotfill & 
20.5500072(6) & 
0.2860686769(4) & 
9.83396979(8) & 
17.3342759(7) & 
3.60511446(5) \\
$x$ (l-s) \dotfill & 
4.466994(6) & 
0.397203(3) & 
9.497573(11) & 
24.39312(8) & 
2.865858(13) \\
$T_0$ (MJD) \dotfill & 
54005.480292(7) & 
54000.1053987(3) & 
54005.80617(6) &
54003.127812(11) & 
54002.7705(11) \\
$e$ \dotfill & 
0.5701606(15) & 
[0.0] & 
0.053108(3) & 
0.712431(2) & 
0.004046(9) \\
$\omega$ ($^\circ$) \dotfill & 314.31935(13) & 0 & 191.500(2) &  131.3547(2) & 323.07(11) \\
$\dot{\omega}$ ($^\circ$ yr$^{-1}$) \dotfill & 0.00391(18) & \nodata & \nodata & 0.00548(30) & \nodata \\
\hline
\multicolumn{5}{l}{Derived parameters} \\
\hline
$f\,(M_{\sun})$ \dotfill & 
0.0002266235(9) & 
0.000822203(16) & 
0.00951180(3) & 
0.0518649(5) & 
0.00194450(3) \\
$M (M_{\sun})$ \dotfill & 2.92(20) & \nodata & \nodata & 1.97(15) & \nodata \\
$M_{p, \rm max}\,(M_{\sun})$ \dotfill & 3.25 & \nodata & \nodata & 1.65 & \nodata \\
$M_{c, \rm min}\,(M_{\sun})$\tablenotemark{a} \dotfill & 0.11 & 0.12 & 0.30 & 0.53 & 0.17 \\
\enddata
\tablenotetext{a}{Calculated assuming a pulsar mass of 1.4
  $M_{\sun}$. Where a $\dot{\omega}$ was measured, these limits
  include 99\% of the total probability.}
\label{tab:binaries}
\end{deluxetable*}

\begin{figure*}[htp]
  \begin{center}
    \includegraphics[width=4.5in,angle=0]{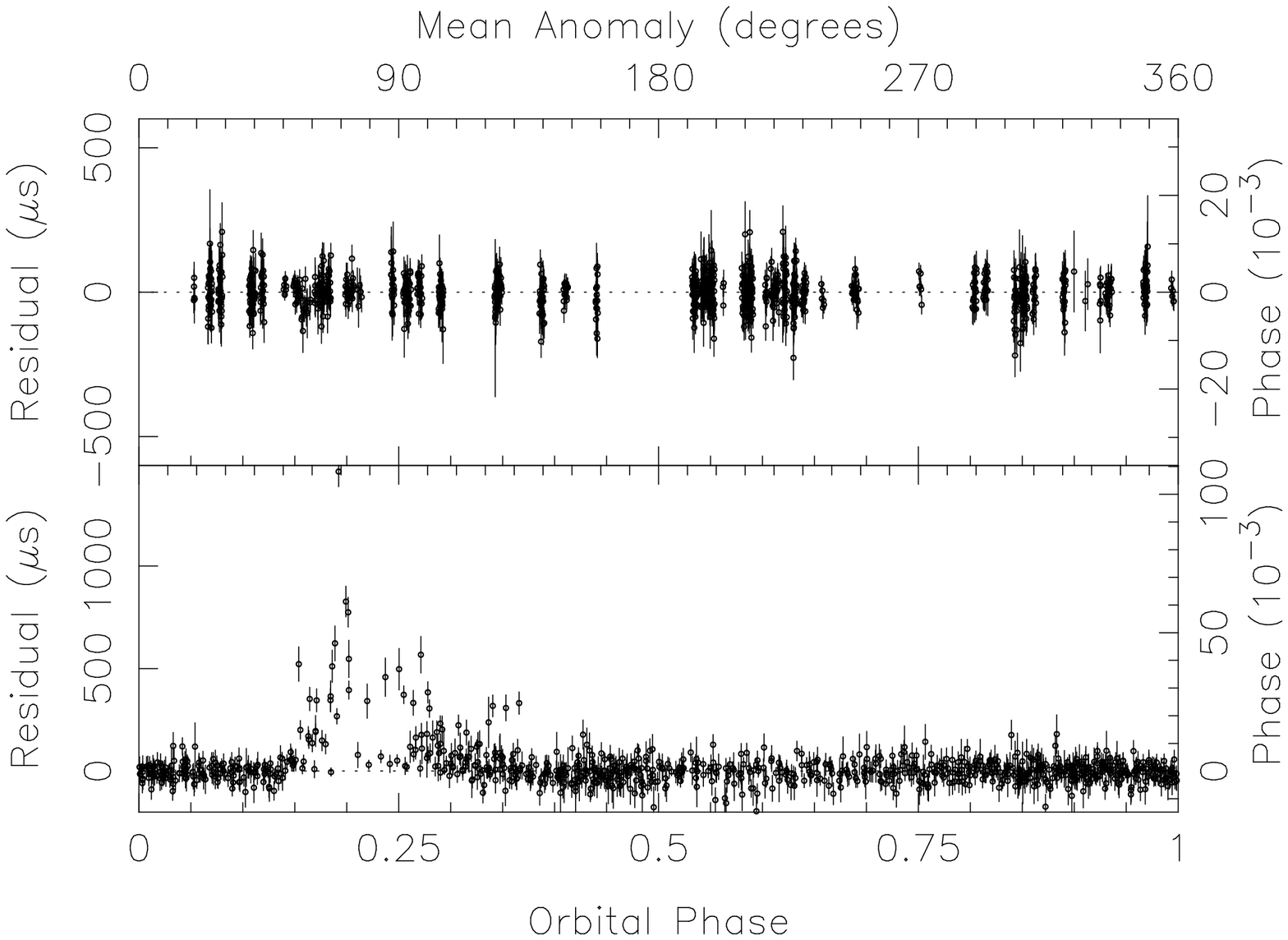}  
    \figcaption{Time of arrival residuals as a function of orbital phase
    for NGC~6440B ({\em top}) and NGC~6440D ({\em bottom}).\label{fig:residuals}}
  \end{center}
\end{figure*}

\section{Timing}
\label{sec:timing}

Diffractive scintillation causes a modulation of the intensity of the
pulsed signal in time and frequency. ``Scintles'' are the regions of
the dynamic spectrum where the signal is amplified relative to
average. The high DMs of these pulsars imply that the scintles have
narrow bandwidths (of the order of $\sim$20 kHz and $\sim$60 kHz at
1.95 GHz for NGC~6440 and NGC~6441, Cordes and Lazio 2001), and are
therefore averaged out over our wide observing band.  This means that
the flux densities of the pulsars are very steady, resulting in the very
high detection rate during our timing campaign: even the faintest new
MSPs are always detectable.  These detections were cross-correlated
with a synthetic profile, obtained from fitting a minimal set of
Gaussian curves to the summed profiles presented in
Figure~\ref{fig:profiles}. This cross-correlation is done in the
Fourier domain, as described in Taylor (1992)\nocite{tay92}, and from
it we derive pulse times of arrival (TOAs). For the faint, isolated
pulsars we calculated about 1 TOA per observation, but for the binary
systems, like NGC~6440B and particularly NGC~6440D, a significantly
larger number were calculated in order to retain orbital phase
information.

These TOAs were fitted to a pulsar model containing spin (its
frequency $\nu$ and its first derivative $\dot{\nu}$) and astrometric
(right ascension $\alpha$ and declination $\delta$) parameters, plus
the DM using {\sc TEMPO}. The parameters obtained from these fits are
presented in Tables~\ref{tab:isolateds} and \ref{tab:binaries}. To
account for the motion of the telescope relative to the barycenter of
the solar system, we used the DE405 Solar System ephemeris
\cite{sta98b}.  The DMs are calculated by comparing pulse TOAs at 820
and 1950 MHz, but they assume that there is no significant
longitudinal evolution of the pulse profile with frequency. For the
binary pulsars, the orbital parameters (orbital period $P_b$;
semi-major axis of the pulsar's orbit $a$ projected along the
line-of-sight in light-seconds, $x \equiv a \sin i/c$; eccentricity
$e$; longitude of periastron $\omega$; time of passage through
periastron $T_0$; and in two cases the rate of advance of periastron
$\dot{\omega}$) were also fitted (see Table \ref{tab:binaries}) using
the the Damour \& Deruelle orbital model \cite{dd85,dd86}. The
reference epoch for all solutions is MJD = 54000 (2006 September 9).
The timing solution of NGC~6440B is the only one that includes the
``mode 14'' data taken in October 2007; all the others include only
mode 2 data, which ends in 2007 March 28.

The 3-level sampling of the Spigot introduces systematics on the
profiles that are not well understood. Following standard pulsar
timing practice in such cases, we compensate for this by increasing
the uncertainties on the TOAs by the small factor indicated in
Tables~\ref{tab:isolateds} and \ref{tab:binaries} as ``EFAC'', such
that the reduced $\chi^2\,=\,1$. In all cases except that of
NGC~6440C, these factors are small, indicating that the timing models
in Tables~\ref{tab:isolateds} and \ref{tab:binaries} do a good job of
predicting the rotational phases of the pulsars, i.e., there are no
unmodeled trends in the TOA residuals (for two examples, see
Fig.~\ref{fig:residuals}). For this reason, we believe that the
1-$\sigma$ uncertainty estimates returned by {\sc TEMPO} are essentially
accurate, so they are reported directly in Tables~\ref{tab:isolateds}
and \ref{tab:binaries}. In \S \ref{sec:NGC6440B}, we test the validity
of this hypothesis for the measurement of the $\dot{\omega}$ of
NGC~6440B.

The large EFAC of NGC~6440C indicates the presence of unmodeled
effects in its TOA residuals. The cause of these is unknown at
present, and will be investigated elsewhere.

\begin{deluxetable}{lr@{.}lr@{.}lr@{.}lr@{.}lr@{.}l}
\scriptsize
\tablecaption{Pulsar offsets from the centers of NGC~6440 and NGC~6441 \label{tab:offsets}}
\tablecolumns{10}
\tablewidth{0pc}
\tablehead{
\colhead{Pulsar} &
\multicolumn{2}{c}{$\theta_{\alpha}$\tablenotemark{a,b}} &
\multicolumn{2}{c}{$\theta_{\delta}$\tablenotemark{a}} &
\multicolumn{4}{c}{$\theta_{\perp}$} &
\multicolumn{2}{c}{$r_{\perp}$} \\
\colhead{} &
\multicolumn{2}{c}{(\arcmin)} &
\multicolumn{2}{c}{(\arcmin)} &
\multicolumn{2}{c}{(\arcmin)} &
\multicolumn{2}{c}{($\theta_c$)} &
\multicolumn{2}{c}{(pc)}
}
\startdata
\multicolumn{10}{l}{NGC 6440}\\
\hline
A \dotfill & $-$0&0017 & $-$0&0447 & 0&04 & 0&34 & 0&11 \\
B \dotfill &    0&0601 & $-$0&0307 & 0&07 & 0&52 & 0&16 \\
C \dotfill & $-$0&3571 & $-$0&2801 & 0&45 & 3&49 & 1&08 \\
D \dotfill & $-$0&2472 &    0&4930 & 0&55 & 4&24 & 1&32 \\
E \dotfill &    0&0240 &    0&1269 & 0&13 & 0&99 & 0&31  \\
F \dotfill & $-$0&0858 & $-$0&0389 & 0&09 & 0&72 & 0&22  \\
\hline
\multicolumn{10}{l}{NGC 6441}\\
\hline
A \dotfill &    0&1797 & $-$0&0993 & 0&21 & 1&87 & 0&81 \\
B \dotfill & $-$0&1432 & $-$0&2989 & 0&33 & 3&01 & 1&30 \\
C \dotfill &    0&1111 & $-$0&0096 & 0&11 & 1&01 & 0&44 \\
D \dotfill &    0&0395 & $-$0&0229 & 0&05 & 0&41 & 0&18 \\
\enddata
\tablenotetext{a}{The uncertainty of these parameters is much smaller
  than the uncertainty of the position of the center of the GC
  (they were computed assuming that the center is {\em exactly} where
  indicated in Table \ref{tab:GC}). This precision is, however,
  necessary to calculate the angular distances between the pulsars.}
\tablenotetext{b}{Note that $\theta_{\alpha} = d \alpha \cos \delta$,
  where $d \alpha$ is the difference of the Right Ascension of the
  pulsar and the center of the GC.}
\end{deluxetable}

\begin{figure}[htp]
  \begin{center}
    \includegraphics[width=3in,angle=0]{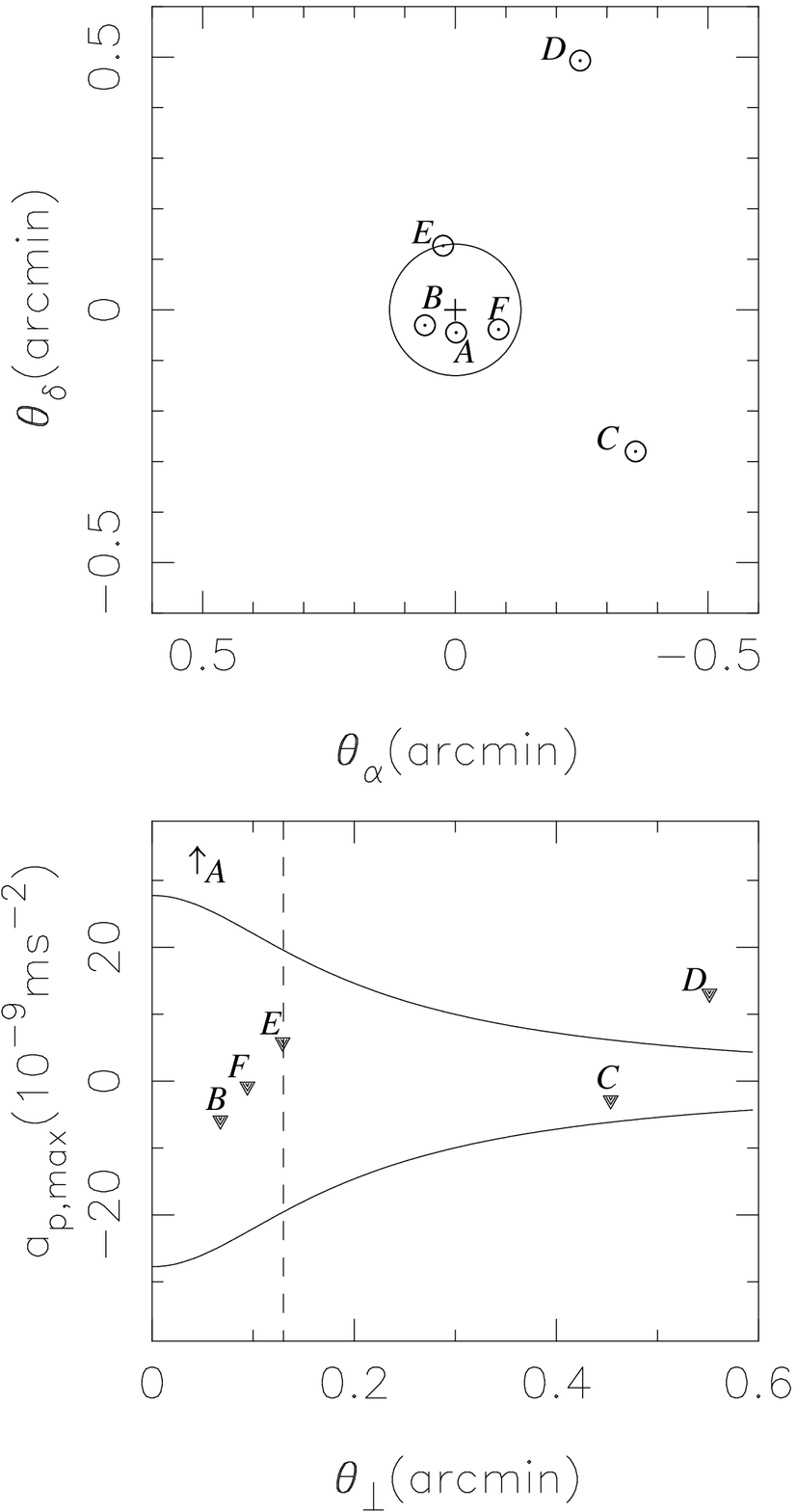}  
    \figcaption{{\em Top}: Positions of the six known pulsars in
    NGC~6440, plotted as west-east ($\theta_{\alpha}$) and south-north
    ($\theta_{\delta}$) offsets from the center of the GC (see also
    Table \ref{tab:offsets}). The angular core radius
    ($\theta_c$ = 0\farcm13) is represented by a circle.
    {\em Bottom}: Upper limits for the accelerations of the
    MSPs ({\em triangles}). These are
    compared with the maximum acceleration that can be caused by the
    GC, as a function of $\theta_{\perp}$ ({\em solid curves}). The
    angular core radius is indicated by the
    vertical dashed line. Pulsar NGC~6440A, and to a much lesser
    extent NGC~6440D are significantly above
    the maximum GC acceleration, indicating that its large
    positive $\dot{P}$ is caused by its intrinsic slowdown.
    \label{fig:NGC6440_offsets}}
  \end{center}
\end{figure}

\begin{figure}[htp]
  \begin{center}
    \includegraphics[width=3in,angle=0]{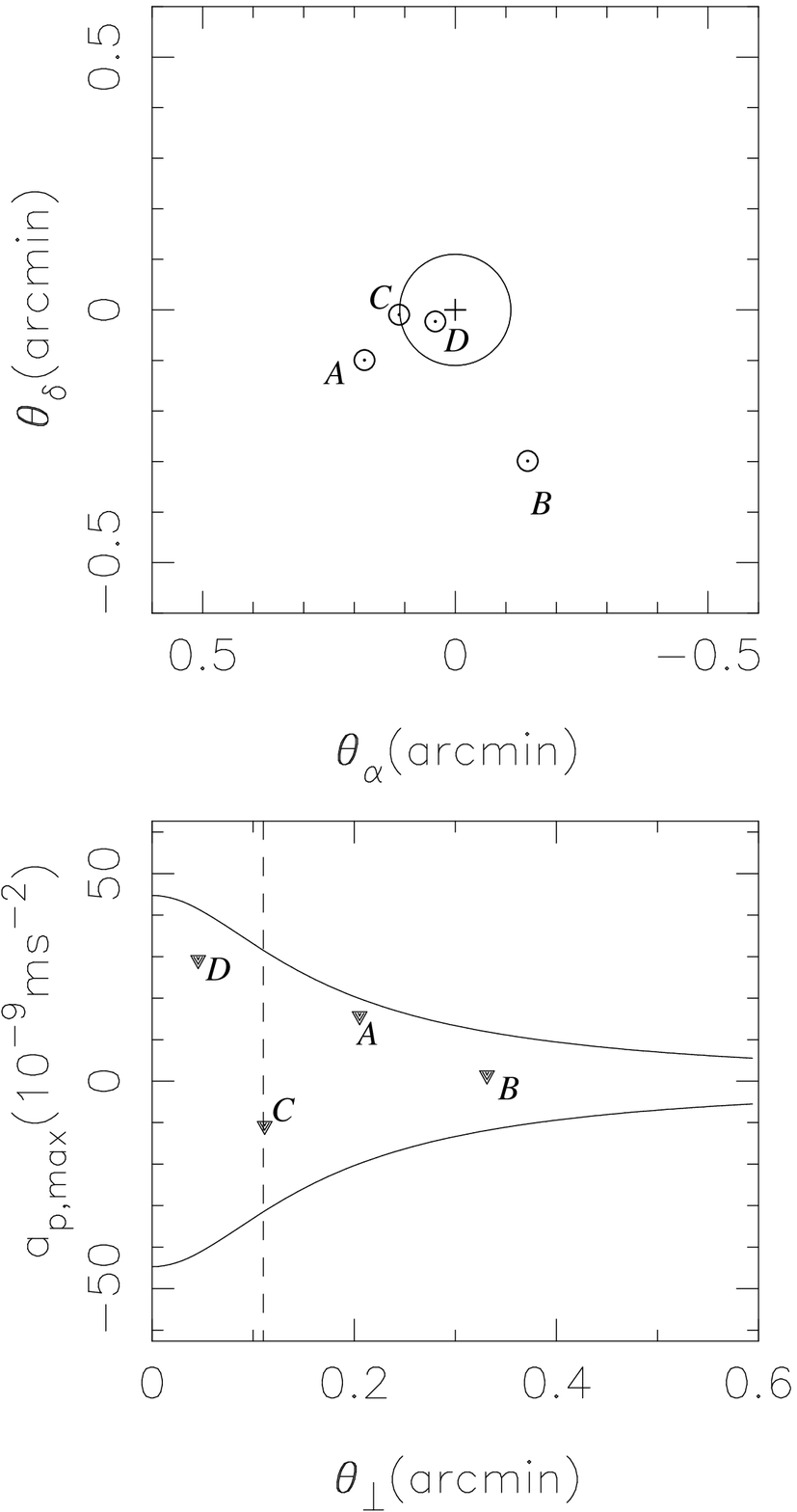}
    \figcaption{{\em Top}: Same as for the top of Fig.~\ref{fig:NGC6440_offsets}
    for the four MSPs in NGC 6441. The angular
    core radius ($\theta_c$ = 0\farcm11) is represented by a circle.
    {\em Bottom}: Upper limits on the accelerations of the MSPs in
    NGC~6441 (triangles) and theoretical maximum acceleration as a
    function of $\theta_{\perp}$ (solid curves).
    \label{fig:NGC6441_offsets}}
  \end{center}
\end{figure}

\subsection{Positions}
\label{sec:positions}

In Table \ref{tab:offsets}, we calculate the angular offsets of the pulsar
positions listed in Tables \ref{tab:isolateds} and \ref{tab:binaries}
relative to the centers of their respective GCs listed in Table
\ref{tab:GC}. These are displayed graphically in the top panels of
Figures \ref{fig:NGC6440_offsets} and \ref{fig:NGC6441_offsets}.
The projected distances of the pulsars from the center of the GC
($r_{\perp}$) were calculated using the most recent GC distance
estimates listed in Table \ref{tab:GC}. There
is nothing unusual about the spatial distribution of the pulsars in
these GCs. As in 47~Tucanae and most other GCs, all pulsars are
located within the half-mass radius of the GC \cite{cr05}; in the
present cases at less than 0\farcm6 (or 4.5 core radii) from the
center. This is not an
observational bias: the S-band telescope beam still has $>\,$50\% sensitivity
$3'$ from the center of the GC. Furthermore, the 820-MHz
observations do not reveal any extra pulsars within $\sim 8\arcmin$ from the
center of the GC. The congregation of pulsars at the center is a
real phenomenon, due to mass segregation in the GC.

\begin{deluxetable*}{lr@{.}lr@{.}lr@{.}lr@{.}lr@{.}lr@{.}lr@{.}lr@{.}lr@{.}lr@{.}l}
\scriptsize

\tablecaption{Limits for derived parameters of the NGC~6440 and NGC~6441 pulsars \label{tab:limits}}
\tablecolumns{21}
\tablewidth{0pc}
\tablehead{
\colhead{Pulsar} &
\multicolumn{2}{c}{$P$} &
\multicolumn{2}{c}{$\dot{P}_{\rm obs}$} &
\multicolumn{2}{c}{$a_{p, \rm max}$} &
\multicolumn{2}{c}{$a_{GC, \rm max}$} &
\multicolumn{2}{c}{$\dot{P}_{\rm int, max}$} &
\multicolumn{2}{c}{$\dot{P}_{\rm int, min}$} &
\multicolumn{2}{c}{$B_{\rm max}$} &
\multicolumn{2}{c}{$B_{\rm min}$} &
\multicolumn{2}{c}{$\tau_{c,\rm max}$} &
\multicolumn{2}{c}{$\tau_{c,\rm min}$} \\
\colhead{} &
\multicolumn{2}{c}{(ms)} &
\multicolumn{2}{c}{(10$^{-18}$)}    &
\multicolumn{4}{c}{($10^{-9}$ m s$^{-2}$)} &
\multicolumn{4}{c}{(10$^{-18}$)}    &
\multicolumn{4}{c}{(10$^9$ G)} &
\multicolumn{4}{c}{(Gyr)}
}
\startdata
\multicolumn{21}{l}{NGC 6440}\\
\hline
A \dotfill & 288&6027917197(16)  &  399&33(10)    &  414&8  & 26&30 & 424&7  & 374&0  & 353&2 & 331&4 & 0&012 & 0&011 \\	
B \dotfill &  16&760127219257(5) & $-$0&32913(16)   & $-$5&86 & 24&72 &   1&05 &  (0&0) &   4&24 &  (0&0) & \multicolumn{2}{l}{($+\infty$)} & 0&25 \\
C \dotfill &   6&226932720487(2) & $-$0&05984(16) & $-$2&85 &  6&15 &   0&07 &  (0&0) &   0&66 &  (0&0) & \multicolumn{2}{l}{($+\infty$)} & 1&45 \\
D \dotfill &  13&495820403909(2) &    0&58678(16) &   13&06 &  4&77 &   0&80 &   0&37 &   3&32 &   2&26 & 0&57 & 0&27 \\
E \dotfill &  16&264003411125(5) &    0&3124(4)   &    5&79 & 19&62 &   1&38 &  (0&0) &   4&78 &  (0&0) & \multicolumn{2}{l}{($+\infty$)} & 0&19 \\
F \dotfill &   3&793629114949(3) & $-$0&01055(18) & $-$0&81 & 22&53 &   0&27 &  (0&0) &   1&03 &  (0&0) & \multicolumn{2}{l}{($+\infty$)} & 0&22 \\
\hline
\multicolumn{21}{l}{NGC 6441}\\
\hline
A \dotfill & 111&6008373133(4)   &    5&661(8)    &   15&67 & 19&88 &  13&2 &  (0&0) &  38&8 &  (0&0) & \multicolumn{2}{l}{($+\infty$)} & 0&13 \\
B \dotfill &   6&074542086503(6) &    0&0192(4)   &    1&41 & 11&88 &   0&27 &  (0&0) &   1&29 &  (0&0) & \multicolumn{2}{l}{($+\infty$)} & 0&36  \\
C \dotfill &  26&56867837497(4)  & $-$0&996(3)   & $-$10&77 & 31&30 &   1&82 &  (0&0) &   7&01 &  (0&0) & \multicolumn{2}{l}{($+\infty$)} & 0&23 \\
D \dotfill &   5&139936774902(4) &    0&4928(2)   &   29&20 & 41&46 &   1&21 &  (0&0) &   2&52 &  (0&0) & \multicolumn{2}{l}{($+\infty$)} & 0&067  \\
\enddata
\end{deluxetable*}

\subsection{Period derivatives}
\label{sec:pdots}

In the third column of Table \ref{tab:limits}, we list the observed
period derivatives ($\dot{P}_{\rm obs}$). Four effects can contribute
to $\dot{P}_{\rm obs}$ \cite{phi93}:

\begin{equation}
\label{eq:pdot}
\left( \frac{\dot{P}}{P} \right)_{\rm obs} =  \left( \frac{\dot{P}}{P}
\right)_{\rm int} + \frac{1}{c} \left( a_{G} + a_{GC} + a_{PM} \right),
\end{equation}

\noindent where $\dot{P}_{\rm int}$ is the pulsar's intrinsic period
derivative, $a_{GC}$ is the line-of-sight component of the
acceleration of the pulsar caused by the gravitational field of the
GC, $a_{G}$ is the difference of the accelerations of the
GC and of the Solar System in the gravitational field of the
Galaxy, projected along the line of sight, $a_{PM} = \mu^2 D$
\citep{shk70} is the centrifugal acceleration due to the transverse
motion of the pulsar $\mu$, and $D$ is the GC's
distance from the Earth, listed in Table~\ref{tab:GC}. The pulsar
proper motions $\mu$ are not yet directly measurable, and there are no
optical measurements of the proper motions of the two GCs; so, like
$\dot{P}_{\rm int}$, $a_{PM}$ is not known but is always positive.

In the fourth column of Table \ref{tab:limits}, we list the
observational upper limits for the pulsar acceleration, $a_{p,\rm
  max}$, obtained by re-arranging all the known terms of
Eq. \ref{eq:pdot} to the right side of the equation:

\begin{equation}
\label{eq:acc}
a_{p,\rm max} \equiv a_{GC} + c \left( \frac{\dot{P}}{P} \right)_{\rm int} + \mu^2 D =
c \left( \frac{\dot{P}}{P} \right)_{\rm obs} - a_G.
\end{equation}

We calculated $a_G$ using a mass model of the Galaxy \cite{kg89}, the
estimated distance to the GCs, and their Galactic coordinates given in
Table~\ref{tab:GC}. Since we have no estimates of the proper motion of
the GCs, we assume from now on that $\mu^2 D = 0$. In all GCs where
this term is measured, it is small relative to the intrinsic period
derivatives and even smaller compared to the accelerations induced by
the GC.

In the fifth column we list the theoretical maximum value of
$a_{GC}$ for each pulsar's line-of-sight, $a_{GC, \rm max}$. This is
calculated using an analytical mass model for the centers of GCs described
in the Appendix of Freire~et~al.~(2005)\nocite{fhn+05}, using the core
radius and central velocity dispersion listed in Table
\ref{tab:GC}. Both $a_{p,\rm max}$ for the individual pulsars and
$a_{GC, \rm max}$ for all lines of sight close to the center of the
GC are displayed in the bottom halves of
Figures~\ref{fig:NGC6440_offsets} and \ref{fig:NGC6441_offsets}.

Using Eq.~\ref{eq:acc}, we can calculate upper (and, in the cases of
NGC~6440A and D, lower)
limits for $\dot{P}_{\rm int}$ assuming the extreme possible
accelerations $\pm a_{GC, \rm max}$. These limits are displayed in the
sixth and seventh columns of Table \ref{tab:limits}. From
these $\dot{P}_{\rm int}$ limits, we can calculate upper and lower
limits for the magnetic field at the surface (estimated using
$B_0\,=\,3.2\,\times\,10^{19}(P \dot{P}_{\rm int})^{1/2}$\,G) and lower and
upper limits for the characteristic age of these pulsars (estimated
using $\tau_c = P / 2 \dot{P}_{\rm int}$), these are displayed in the final
columns of Table \ref{tab:limits}. No pulsar in either GC has
$a_{p,max} < -a_{GC, \rm max}$, this means that the present
GC mass models can predict line-of-sight accelerations large
enough to explain the observed negative period derivatives (and
plausibly maintain that the pulsars are not spinning up).

\subsubsection{NGC~6440A}
\label{sec:NGC6440A}

As previously found by Lyne, Manchester \& D'Amico
(1996)\nocite{lmd96}, PSR~B1745$-$20A has a characteristic age
$\sim 10^3$ times smaller than the age of the GC and
$B_0\,\sim\,3\,\times\,10^{11}$G, a value that is more typical of
what one finds in the general Galactic population. For these two
reasons, there have been some doubts about the association of this
pulsar with NGC~6440. The discovery of five new MSPs in NGC~6440 at
DMs similar to that of PSR~B1745$-$20A confirms its association with
NGC~6440. Furthermore, PSR~B1745$-$20A is closer to the center of the
GC ($\theta_\perp \,=\,0\farcm04$) than any of the new MSPs.

This association of a ``young'' pulsar with a GC is not unique to
NGC~6440. The first known examples were PSR~B1718$-$19, in NGC~6342
\cite{lbhb93} and PSR~B1820$-$30B in NGC~6624 \cite{bbl+94}. The
possible origins of these objects, and in particular NGC~6440A, are
discussed in detail in Lyne, Manchester \& D'Amico (1996). Either
neutron star formation continues to happen in GCs ---
through electron capture supernovae (ECS), which can form when an
accreting WD star reaches the Chandrasekhar limit or when a WD binary
coalesces --- or some form of mild recycling happened, in which the
pulsar was not spun up to very short spin periods and not much of its
magnetic field was buried. The ECS scenario is particularly appealing
as this type of supernova also produces small kicks, which nicely
solves the problem of neutron star retention in GCs
(Ivanova et al. 2007)\nocite{ihr+07}.

\subsubsection{NGC~6440D}

Unlike in the case of NGC~6440A, the upper limit for $\tau_c$
and the lower limit for $B$ does not
imply that this pulsar is younger or has a stronger magnetic field
than the other MSPs. It is possible that the other ``slow MSPs''
(with $13\,<\,P\,<\,27\,$ms) have similar ages and magnetic fields as
those of NGC~6440D; their intrinsic $\dot{P}$s might
simply being masked by the GC accelerations. The apparently special
limits for NGC~6440D might have more to do with its large distance
from the center of the GC: the range of possible GC
accelerations $a_{GC, \rm max}$ is the smallest for all pulsar
positions in NGC~6440 and NGC~6441; this leads to tighter limits on
$\dot{P}_{\rm int}$. If NGC~6440D is a typical slow MSP, then the
magnetic fields of these objects are one order of magnitude
larger than those observed in 47~Tucanae \cite{fcl+01}; this would
cause the observed difference in spin periods between GCs discussed in
\S \ref{sec:periods}.

\section{Binary systems}
\label{sec:binary}

Before this survey, only one binary pulsar was known in these two GCs,
NGC~6441A \cite{pcm+06}. We have discovered four new binary systems,
three in NGC 6440 (B, D and F) and one in NGC~6441 (B). These binaries
make up half of the total observed pulsar population in these GCs.
This ratio is similar to that of Terzan~5, but significantly lower
than what we find in 47~Tucanae and M62, in the latter GC all the 6
known pulsars are in binary systems.

\subsection{NGC 6440F and NGC 6441B}

These two binaries, with orbital periods of 9.8 and 3.6\,d and minimum
companion masses of 0.30 and 0.17~$M_{\sun}$ respectively (assuming
pulsar masses of $1.4\,M_{\sun}$),
might be the only ``normal'' MSP-WD binaries among the new
pulsars. Their spin periods (3.79 and 6.07 ms) are at the lower end of
the spin period distribution found in NGC~6440 and NGC~6441, but they
are typical of what is found among similar MSP-WD binaries in other
GCs and in the Galactic disk. Their eccentricities (0.0531 and
0.00404) are small
compared to the highly eccentric NGC~6440B and NGC~6441A, but they are
nevertheless much larger than what is found in binary systems with
similar orbital periods in the Galactic disk. This is a common
occurrence in GCs, and is probably due to perturbations caused by
close flybys of other stars in these GCs \cite{hr96}.

\subsection{NGC 6440D - Eclipsing binary}

The binary system NGC~6440D has an orbit with non-measurable
eccentricity and a period of 6.9\,hr. It is an eclipsing system with
eclipses that usually last for about 10\% of the orbital cycle. Its
minimum companion mass is $0.12\,M_{\odot}$, calculated assuming a
pulsar mass of $1.4\,M_{\odot}$.  Freire~(2005)\nocite{fre05} called
attention to the fact that eclipsing binary pulsars have a bi-modal
distribution of mass functions: those with $f\,>\,10^{-4}\,M_{\sun}$
were designated ``Eclipsing Low-Mass Binary Pulsars'' (ELMBPs), while
those with $f\,<\,3 \times 10^{-5}\,M_{\sun}$ were designated as
``Very Low-Mass Binary Pulsars'' (VLMBPs).  The VLMBPs are more
commonly known as ``Black Widow'' pulsars.  NGC~6440D is clearly a
ELMBP.  Unlike ``Black Widows'', these objects only occur in GCs,
suggesting that they form through exchange interactions when a radio
pulsar acquires a MS companion.

Sometimes NGC~6440D is detectable during superior conjunction. This
indicates that the orbital inclination is certainly less than
90$^\circ$, otherwise the companion itself would cause an eclipse. It
also implies that the concentration of the material producing the
eclipse is highly variable with time.  Nevertheless, the cloud of
material seems to generally lead the companion in orbital phase (see
Fig.~\ref{fig:residuals}), i.e., ingress occurs at a much larger
distance from superior conjunction than egress. Before, after, and
sometimes through the eclipse we detect $\sim$0.5\,ms delays in the
times of arrival (see Fig.~\ref{fig:residuals}). These are probably
due to an increase in the electron column density near the companion
of up to $\sim$0.5\,cm$^{-3}$\,pc (or about $1.5 \times 10^{18} \rm
cm^{-2}$).  This is of the same order of magnitude as what has been
found in similar binary systems, e.g. M30A \cite{rsb+04}.

If the system is nearly edge on, with pulsar and companion masses of
$1.4$ and $0.124\,M_{\sun}$, the separation between the stellar
components is 1.47$\times$10$^{11}$\,cm ($2.1\,R_{\sun}$). On some
occasions the eclipse extends up to $\sim 30^\circ$ from superior
conjunction (normally before the companion), which implies the
detection of material at $\sim 1.0\,R_{\sun}$ from the center of the
companion.  The size of the Roche lobe of the companion is
$0.42\,R_{\sun}$, therefore, it is clear that we are detecting the
presence of material that is not bound to the companion at ingress. We
almost never detect unbound material at egress.

This is different from what one observes in ELMBPs such as 47~Tuc~W
\cite{clf+00}, Terzan~5~P and ad \cite{rhs+05,hrs+06},
PSR~J1740$-$5340 in NGC~6397 \cite{dpm+01} or M28~H (B\'egin 2006;
B\'egin et al. in preparation)\nocite{beg06,brf+08}, which display
long duration irregular eclipses likely related to excessive unbound
ionized material from the companion over much of the orbit.  Three of
these ELMBP companions have been identified at optical wavelengths
(47~Tuc~W, Edmonds et al. 2002, \nocite{egc+02}; PSR~B1718$-$19 in
NGC~6349, Kerkwijk et al. 2000\nocite{kkk+00}; and PSR~J1740$-$5340 in
NGC~6397, Ferraro et al. 2001 \nocite{fpds01}), and in all cases the
observed companion is a low-mass, non-degenerate star.  None of the
ELMBPs with shorter eclipses such as M30~A, Terzan~5~A \cite{ljm+90}
and M62~B \cite{pdm+03} have been optically detected.  The situation
is similar at X-rays: only the systems with large eclipses have been
detected (e.g., 47~Tuc~W, see Bogdanov, Grindlay \& van den Berg
2006)\nocite{bgb05}.  In these systems the hard X-ray emission is
generated by the collision of MSP and stellar winds.  None of the
ELMBPs with shorter eclipses have been definitively detected at X-ray
wavelengths.  If NGC~6440D is similar to these ELMBPs, it seems
less likely that it will be detected at either optical or X-ray
wavelengths.

When an exchange encounter forms a ELMBP, the ejection of the previous
companion to the pulsar imparts a kick to the binary, making it fly
away from the central regions of the GC in an eccentric orbit. It is
probably for that reason that two ELMBPs (PSR~B1718$-$19 and
PSR~J1740$-$5340) are found at large distances from the centers of
their parent GCs \cite{fre05}. NGC~6440D follows this trend: of all
the pulsars in NGC~6440 and NGC~6441, it is farthest from the center of
its GC as projected on the plane of the sky.

\subsection{NGC 6440B: eccentric binary with a super-massive neutron star?}
\label{sec:NGC6440B}

NGC~6440B and NGC~6441A (\S \ref{sec:NGC6441A}) are part of a diverse
class of binary pulsars in GCs with eccentric orbits that includes
mostly recent discoveries like M30B \cite{rsb+04}, NGC~1851A (Freire
et al.~2004; Freire et al. 2007)\nocite{fgri04,frg07}, six systems in
Terzan~5 (I, J, Q, U, X and Z; Ransom et al.~2005; Stairs et al., in
preparation)\nocite{rhs+05} and M28~C and D (B\'egin 2006; B\'egin et
al. in preparation)\nocite{beg06,brf+08}. With the exception of M15C
\cite{jcj+06}, which is very similar to some DNSs seen in the Galaxy,
they have no Galactic analogues.  All of these systems were likely
formed through stellar exchange encounters.

The orbital eccentricity of NGC~6440B has allowed a highly significant
measurement of the rate of advance of periastron:
$\dot{\omega}\,=\,0.00391(18)^\circ \rm yr^{-1}$.
As mentioned above, this is a 1-$\sigma$ uncertainty derived by {\sc
TEMPO}. To check if this uncertainty value is realistic, we estimated
$\dot{\omega}$ and its associated uncertainty using two other
methods. First, all the timing parameters of NGC~6440B were estimated
using a Monte-Carlo Bootstrap algorithm \cite{et93,ptvf92}, with a
total of 1024 fake TOA datasets that are consistent with the real
dataset. From this we obtain $\dot{\omega}\,=\,0.00391(16)^\circ \rm yr^{-1}$.
In the second estimate, we kept $\dot{\omega}$ fixed and fitted
all the remaining timing parameters, recording the resulting
$\chi^2$. Doing this for a range of values of $\dot{\omega}$, we can
estimate the 1-$\sigma$ uncertainty as the half-width of the region
where $[\chi^2 (\dot{\omega}) - \chi^2 (\dot{\omega}_{\rm min}) < 1]$, being
$\dot{\omega}_{\rm min}$ the value that minimizes $\chi^2$ \cite{sna+02}.
The result of this estimate is $\dot{\omega}\,=\,0.00391(18)^\circ \rm
yr^{-1}$. Both these estimates indicate that, at least for the
$\dot{\omega}$ of NGC~6440B, the 1-$\sigma$ {\sc TEMPO} uncertainty is
reliable.

The orbital coverage of the timing data is excellent (see
Fig.~\ref{fig:residuals}): the orbit was well sampled at
the end of 2005, at the start of 2007 and again during October
2007. For this reason, $\dot{\omega}$ is not strongly correlated to
any timing parameter: its largest correlation is with $P_b$ (96\%),
which is normal for this particular parameter.

\begin{figure*}[htp]
  \begin{center}
    \includegraphics[width=5in,angle=0]{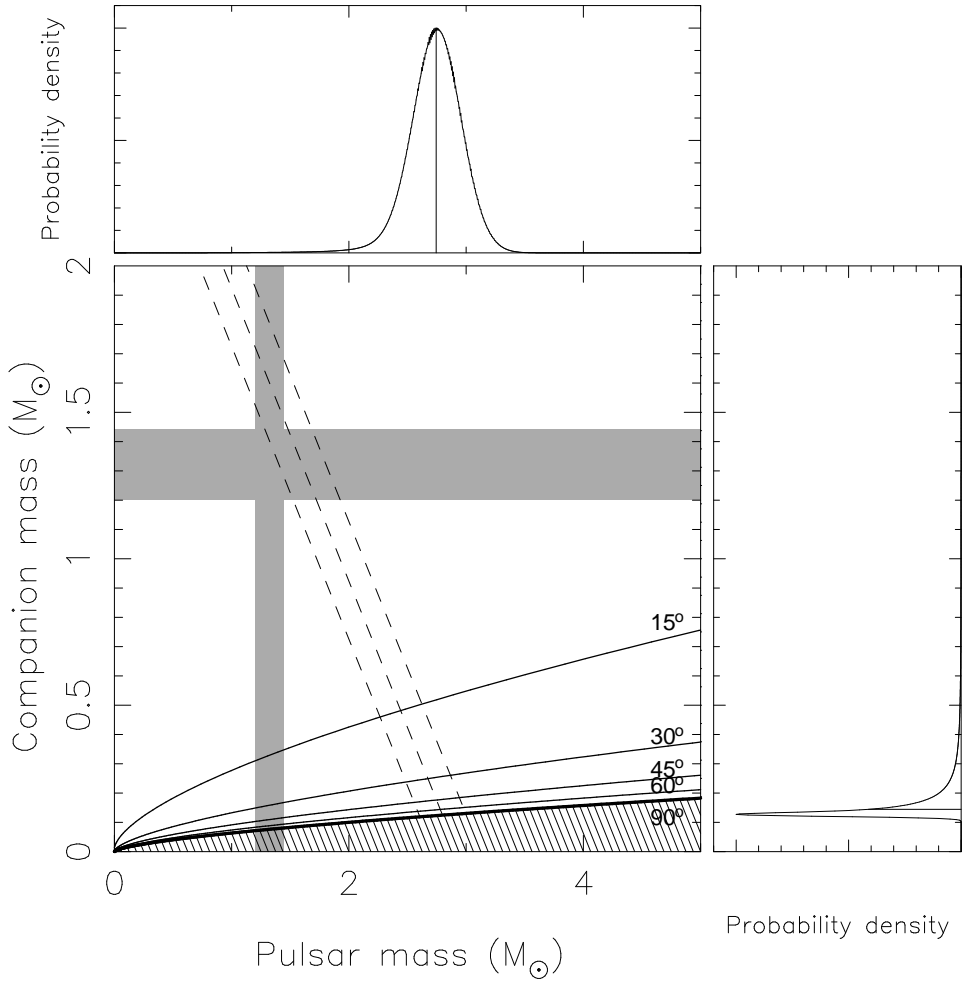}
    \figcaption{Graphical display of the constraints on the masses of
    NGC~6440B and its companion. In the main display, the hatched
    region is excluded by knowledge of the mass function and by $\sin
    i \leq 1$. The slanting straight lines correspond to a total
    system mass that causes a general-relativistic $\dot{\omega}$
    equal or within 1\,$\sigma$ ({\em dashed lines}) of the measured
    value. The five solid curves indicate constant orbital inclinations. The
    gray bars indicates the range of precisely measured neutron star
    masses.
    We also display the probability density function for the
    mass of the pulsar ({\em top}) and the mass of the companion ({\em
    right}), and mark the respective medians with vertical
    (horizontal) lines.
    \label{fig:NGC6440B_mass_mass}}
  \end{center}
\end{figure*}

\subsubsection{Mass estimate from periastron advance}

Assuming that $\dot{\omega}$ is fully relativistic (an important
assumption, as we will see) implies a total system mass of
$2.92\,\pm\,0.20\,M_{\sun}$. This binary system is located, at least
in projection, close to the center of the GC, where the massive
objects should predominantly occur. In Figure~\ref{fig:NGC6440B_mass_mass},
we present the mass constraints on the components graphically. We
assume that the probability density function (pdf) for $\dot{\omega}$,
$p(\dot{\omega})$, is a Gaussian, with half-width similar to the
1$\,\sigma$ uncertainty given by {\sc TEMPO}. We also assume a constant
probability density for $\cos i$, where $i$ is the orbital inclination
($90^\circ$ for an edge-on orbit). From these constraints, we
calculate a two-dimensional pdf for
the mass of the pulsar and the mass of the companion. This is then
projected in both dimensions, resulting in final pdfs for the mass of
the pulsar and the mass of the companion.

\begin{figure*}[htp]
  \begin{center}
    \includegraphics[width=5in,angle=0]{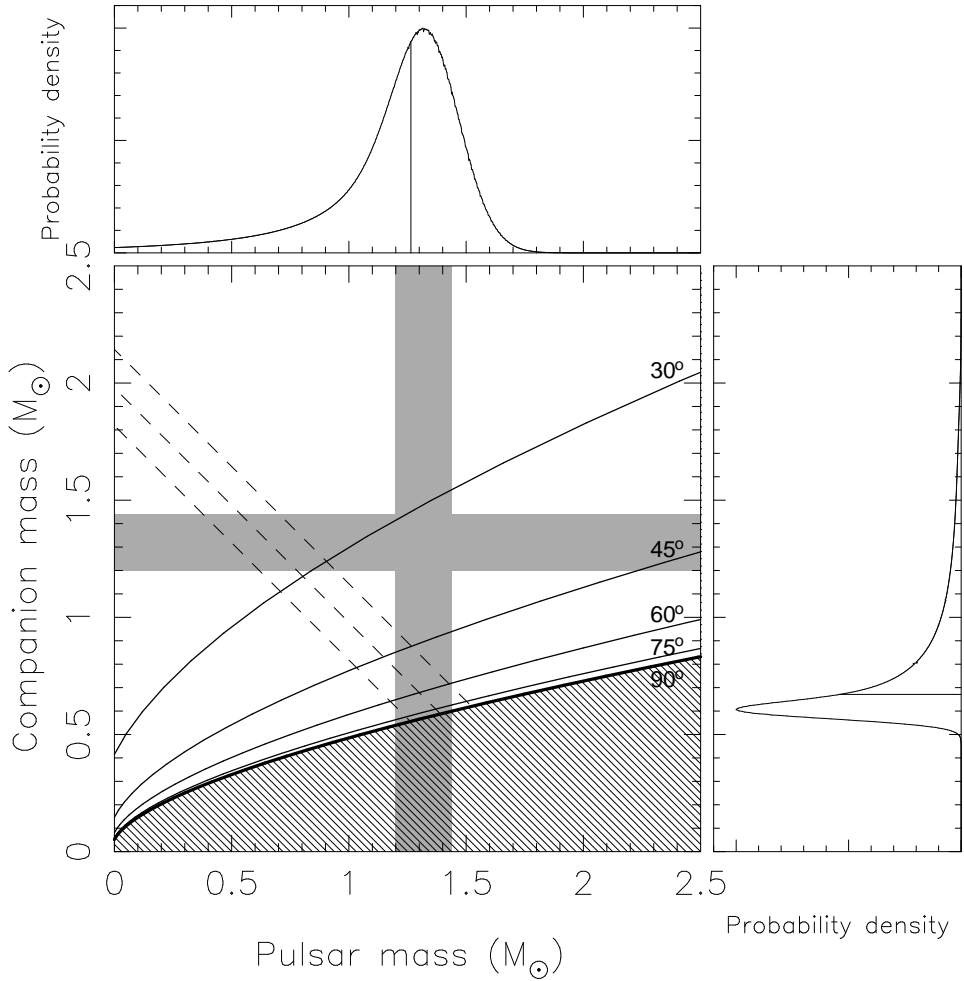}
    \figcaption{Display as in Fig.~\ref{fig:NGC6440B_mass_mass}, this
    time for the NGC~6441A binary system.
    \label{fig:NGC6441A_mass_mass}}
  \end{center}
\end{figure*}

The total mass of this system suggests that it is a double neutron
star (DNS) binary. However, the probability that the pulsar lies
within the range of neutron star masses that have been precisely
measured to date --- from $\sim 1.20\,M_{\sun}$ for the companion of
PSR~J1756$-$2251 \cite{fkl+05} to $1.44\,M_{\sun}$ for PSR~B1913+16
\cite{wt03}, indicated in Figures~\ref{fig:NGC6440B_mass_mass} and
\ref{fig:NGC6441A_mass_mass} by grey bars --- is only 0.10\%. This
eventuality would require very low orbital inclinations, between $\sim
4^\circ$ and $\sim 5^\circ$. In this inclination range, it is also
possible that the pulsar has a blue straggler companion, a possibility
that can be investigated using optical/IR imaging.

Before this study, the most massive neutron star known was
PSR~B1516+02B in the GC M5 ($1.96^{+0.09}_{-0.12}
M_{\sun}$; Freire et al. 2007)\nocite{fwb+07}. If NGC~6440B had a
similar mass, then the companion could be a $\sim
0.9\,M_{\sun}$ white dwarf, or a blue straggler of similar mass.
However, we note that there is only a 0.97\%
probability that the mass of the pulsar is below $2.0\,M_{\sun}$.

The median pulsar mass is 2.74~$M_{\sun}$, with the lower and upper
1\,$\sigma$ limits at 2.52 and 2.95~$M_{\sun}$, and the
2\,$\sigma$ limits at 2.23 and 3.15~$M_{\sun}$.  The pulsar has a 99\%
probability of being less massive than 3.24~$M_{\sun}$. For the
companion, the median of the distribution is at $0.142\,M_{\odot}$,
with the the lower and upper 1\,$\sigma$ limits at $0.124$ and
$0.228\,M_{\odot}$, and the 2\,$\sigma$ limits at $0.113$ and
$0.571\,M_{\odot}$, implying that it could be a low-mass WD or an
un-evolved MS star. If this high pulsar mass is confirmed, it would be
by far the largest neutron star mass ever measured. That would have
profound consequences for the study of the equation of state of dense
matter, since almost no models predict neutron stars more massive than
2.5~$M_{\sun}$ \cite{lp07}.

If NGC~6440B really has a mass of $\sim 2.7\, M_{\odot}$, it would
imply that the end products of the coalescence of DNS systems might
themselves be stable as super-massive neutron stars.  The total masses
of the known DNS systems range from $2.57\,M_{\sun}$ \cite{fkl+05} to
$2.83\,M_{\odot}$ \cite{wt03}. Such coalescence products might be
observable after coalescence, either through gravitational wave
emission \cite{and03}, cooling through neutrino emission (direct URCA
process, Page \& Applegate 1992\nocite{pa92}) or even high-energy
blackbody emission.

It is possible that NGC~6440B itself formed this way. M15C, the only
DNS known in the GC system \cite{jcj+06}, is expected to coalesce in
about $3 \times 10^8$ years ($\sim 3$\% of the age of M15).  If the
products of DNS coalescence are as likely to be detected as radio
pulsars as their progenitors, then we should detect $\sim 30$ of them
in the whole GC system. It is possible that some isolated MSPs in GCs
and the Galactic disk formed this way. The massive isolated MSPs in
GCs can later acquire a companion through exchange encounters, forming
a system like NGC~6440B where we can measure their large masses.

\subsubsection{Is the periastron advance relativistic?}

The previous discussion rests on the assumption that the observed
$\dot{\omega}$ is fully relativistic. If there was an extra
contribution to $\dot{\omega}$ from tidal or rotational deformation of
the companion, then no reliable estimates can be made of the total
mass of the system.

Lai, Bildsten \& Kaspi (1995)\nocite{lbk95}, in their analysis of
the binary pulsar PSR~J0045$-$7319, concluded that the only likely
contribution to  $\dot{\omega}$ in that system is from rotational
deformation of the companion. NGC~6440B has a shorter orbital period
than  PSR~J0045$-$7319 (20.5 versus 51 days), but it also has a
much smaller companion mass ($\sim 0.12\, M_{\sun}$ at $i = 90^\circ$
to $\sim 1.5\,M_{\sun}$ at $i \sim 4-5^\circ$)
versus $\sim 9\, M_{\sun}$ for the companion of PSR~J0045$-$7319
\cite{bbs+95}. Main-sequence stars of such masses have
radii of $R \sim 0.18 - 1.5$ and $\sim 4.5\,R_{\sun}$ respectively
\cite{lang91}. This implies that even if the companion of NGC~6440B
were extended, the contribution to $\dot{\omega}$ from its tidal
deformation (the latter proportional to $(R/a)^3$, where $a$ is the
separation between components) would always be
smaller for NGC~6440B than it is for PSR~J0045$-$7319.

The only possible exception to this is if the companion of NGC~6440B
were a $\sim 0.9 \, M_{\sun}$ star undergoing a pre-giant or giant
phase. Given the mass function of the system, that would require a
very low orbital inclination. A giant star of that mass would have a
radius of up to 1 a.u. \cite{lang91}, and would not fit in the space
between the pulsar and companion at any orbital phase. If it were in a
``pre-giant'' phase, its atmosphere would be very extended and have
significant mass loss, leading to variations in the DM with orbital
phase. The resulting tides should not only have circularized the
orbit, but they should also have directly observable effects on the
timing: unmodeled effects in the rotation of the pulsar and random
variations of the orbit's period and apparent size (Nice, Arzoumanian
\& Thorsett 2000)\nocite{nat00}. For PSR~J0045$-$7319, there is strong
orbital decay due to tidal effects \cite{kq98}.

For NGC~6440B, we obtain $\dot{P_b}\,=\,(-0.0 \pm
1.4)\,\times\,10^{-9}$ and
$\dot{x}\,=\,(-0.21\,\pm\,0.14)\,\times\,10^{-12}$. The pulsar times
very well, with an EFAC similar to those of the isolated MSPs in the
GC (see Table \ref{tab:binaries}), no unmodeled systematic trends
are visible in the TOA residuals (Fig.~\ref{fig:residuals}). Fitting
for the halves of the orbit ``in front'' and ``behind'' the companion,
we see no DM variations larger than our detection limit of 0.07
cm$^{-3}$pc. The possibility of an unusually extended companion to
NGC~6440B and of a significant tidal contribution to $\dot{\omega}$
can be safely excluded.

The contribution from rotational deformation, henceforth designated as
$\dot{\omega}_{\rm{rot}}$, could be significant if the companion
(degenerate or not) were spinning rapidly. The rotational deformation
will also induce a change in the projected semi-major axis of the
orbit $\dot{x}$. Splaver~et~al.~(2002)\nocite{sna+02}, based on the
results of Lai, Bildsten \& Kaspi (1995) and Wex~(1998)\nocite{wex98},
relate $\dot{\omega}_{\rm{rot}}$ to $\dot{x}$:
\begin{equation}
\dot{\omega}_{\rm rot} = 
  \frac{\dot{x}}{x}\,
  \left(
  \tan i
    \frac{1-\frac{3}{2}\sin^2\theta}{%
          \sin\theta\cos\theta\sin\Phi_0}
    + \cot \Phi_0
  \right),
\end{equation}
where $i$ is the orbital inclination of the binary, $\theta$ is the
angle between the angular momentum vector of the secondary and the
angular momentum vector of the orbit; and $\Phi_0$ is the longitude of
the ascending node in a reference frame defined by the total angular
momentum vector (see Fig.~9 of Wex 1998).

Thus, our observed upper limit of  $|\dot{x}/x|< 8 \times 10^{-14}\,$
implies $| \dot{\omega}_{\rm{rot}}|\,<\,1.5 \, \times
\,10^{-4}\,^{\circ}$\,yr$^{-1}$ times a geometric factor. In 80\% of
cases this geometric factor will be smaller than 10. Therefore
$\dot{\omega}_{\rm{rot}}$ will in most cases be of the same order of
magnitude as the present measurement uncertainty for $\dot{\omega}$.
Certain special combinations of $i$, $\theta$, and $\Phi_0$
will make the geometric factor large, but, again, this would only
occur if the companion were rotating fast.

To summarize, the contribution to the $\dot{\omega}$ of this system
from the tidal deformation of the companion is negligible.
Furthermore, observational limits on $\dot{x}$ indicate that
$\dot{\omega}_{\rm{rot}}$ is also likely to be negligible and
the observed $\dot{\omega}$ due to the effects of general
relativity.

\subsection{NGC 6441A: eccentric binary with a massive companion}
\label{sec:NGC6441A}

NGC~6441A was the only pulsar previously known in NGC~6441
\cite{pcm+06}, although its phase-connected timing solution has never
been published. Its orbital period is 17.3 days, and its orbit is
highly eccentric ($e\,=\,0.71$), allowing a measurement of the advance
of periastron for this system, $\dot{\omega}\,=\,0.00548(30)^\circ \rm
yr^{-1}$.  Assuming, again, that this effect is fully due to general
relativity, we obtain a total mass of $1.97\,\pm\,0.15\,M_{\sun}$,
consistent with the estimate made by Possenti et al. (2006):
$2.15\,\pm\,0.06\,M_{\sun}$. Using the methods applied in the case of
NGC~6440B, we calculated pdfs for the mass of this pulsar and its
companion for our measurement of the total mass of the binary; these
are displayed graphically in Figure~\ref{fig:NGC6441A_mass_mass}.
There is a 99\% probability that the companion is more massive than
0.53\,$M_{\sun}$ and that the pulsar is less massive than
1.65\,$M_{\sun}$. The medians for the pulsar and companion masses are
1.26 and 0.67\,$M_{\sun}$ respectively, and there is a 45.6\%
probability that the pulsar mass is within the 1.20-1.44~$M_{\sun}$
mass range.  The nature of the companion is unknown, but no
observational hints of tidal effects ($\dot{x}$, $\dot{P_b}$, DM
variations) are present, suggesting it is a compact object.

\section{Conclusion and prospects}

We have discovered eight new pulsars in the GCs NGC~6440
and NGC~6441. Their pseudo-luminosities indicate that there may be as
many pulsars in these GCs as in Terzan~5, but less pulsars are
observed in the former because of their larger distances. The pulsar
population in these GCs seems to have distinctly lower spin
frequencies. To some extent, this is due to the presence of
apparently young objects like NGC~6440A. At least in the case of
NGC~6440A and D, and possibly in the case of the other slow MSPs, the
larger spin periods could be related with many of them having
higher magnetic fields than the MSPs in other GCs.

Four of the new pulsars are in binary systems. NGC~6440D is an
eclipsing binary, with a companion star with a mass of about
0.12~$M_{\sun}$ or larger. NGC~6440B is a 16.7-ms pulsar in a 20.5-day
orbit with an orbital eccentricity of 0.57. A measurement of the rate
of advance of periastron for this system suggests that this binary
hosts the most massive neutron star to date:
$2.74\,\pm\,0.21\,M_{\sun}$, with only a 1\% probability that the
inclination is low enough that pulsar mass is below $2\,M_{\sun}$.
This result depends on the (likely) possibility that the observed
periastron advance is fully relativistic. Such a large mass would introduce
the strongest constraints to date on the equation of state; it also
suggests that the products of the coalescence of double neutron stars
might be stable neutron stars. The rate of advance of
periastron was also measured for NGC~6441A, indicating a significantly
less massive pulsar ($M_p\,<\,1.65\,M_{\sun}$).

The pulsars in NGC~6440 will be monitored carefully over the next few
years. In the case of NGC~6440C, the new data will allow a detailed
investigation of the origin of its timing irregularities. In the case
of NGC~6440B, the continued monitoring will lead to an improvement in
the precision of $\dot{\omega}$. Simulations indicate that the
Einstein delay ($\gamma$) may become measurable with about 15 years of
data; this would provide unambiguous mass measurements and determine
the nature of the companion. Deep optical studies of NGC~6440 might be
extremely valuable in determining the nature of the companion of
NGC~6440B.

\acknowledgements

The National Radio Astronomy Observatory is a facility of the National
Science Foundation operated under cooperative agreement by Associated
Universities, Incorporated. IHS held an NSERC UFA while most of
this work was performed and pulsar research at UBC is supported by a
Discovery Grant. JWTH holds an NSERC PDF and CSA
supplement. LHF's work for this paper was sponsored by the Research
Experience for Undergraduates program of the NSF.

\bibliographystyle{../apj1d}

\typeout{get arXiv to do 4 passes: Label(s) may have changed. Rerun}
\end{document}